\documentclass{ecai}
\usepackage{graphicx}
\usepackage{latexsym}

\usepackage{marvosym}
\usepackage{fontawesome}
\usepackage{amsmath}
\usepackage{amssymb}
\usepackage{amsthm}
\usepackage{arydshln}
\usepackage{listings}
\usepackage{subcaption}
\usepackage{multirow}
\usepackage{multicol}
\usepackage{tikz}
\usetikzlibrary{arrows,automata}
\usetikzlibrary{shapes}
\newcommand{\tup}[1]{{\langle #1 \rangle}}
\usepackage{xcolor}
\usepackage{todonotes}
\usepackage{enumitem}
\usepackage{mathtools}

\usepackage{listings}
\newcommand{\strips}{\textsc{Strips}}     

\begin{document}

\begin{frontmatter}

\title{Synthesis of Procedural Models for Deterministic Transition Systems}

\author[A]{\fnms{Javier}~\snm{Segovia-Aguas}\thanks{Corresponding Author. Email: javier.segovia@upf.edu.}}
\author[B]{\fnms{Jonathan}~\snm{Ferrer-Mestres}}
\author[C]{\fnms{Sergio}~\snm{Jim\'enez}}

\address[A]{Universitat Pompeu Fabra}
\address[B]{CSIRO}
\address[C]{Universitat Polit\`ecnica de Val\`encia}

\begin{abstract}
This paper introduces a general approach for synthesizing procedural models of the {\em state-transitions} of a given {\em discrete system}. 
The approach is general in that it accepts different {\em target languages} for modeling the state-transitions of a discrete system; different model acquisition tasks with different target languages, such as the synthesis of \strips{} action models, or the update rule of a {\em cellular automaton}, fit as particular instances of our general approach. We follow an inductive approach to synthesis meaning that a set of examples of state-transitions, represented as {\em (pre-state, action, post-state)} tuples, are given as input. The goal is to synthesize a structured program that, when executed on a given pre-state, outputs its associated post-state. Our synthesis method implements a combinatorial search in the space of well-structured terminating programs that can be built using a {\em Random-Access Machine} (RAM), with a minimalist instruction set, and a finite amount of memory. The combinatorial search is guided with functions that asses the  complexity of the candidate programs, as well as their fitness to the given input set of examples.   
\end{abstract}

\end{frontmatter}

\section{Introduction}
Transition systems make it possible to describe the behavior of discrete systems and white-box models of such systems are used in AI with  different purposes; to name a few, the building of {\em simulators}~\cite{fox1988knowledge,fishwick2012knowledge}, {\em causal inference}~\cite{pearl2018book}, {\em automated planning}~\cite{ghallab2004automated,geffner2013concise}, {\em model-checking}~\cite{clarke1997model}, {\em system diagnosis}~\cite{reiter1987theory}, {\em epistemic reasoning}~\cite{fagin2004reasoning}, or {\em game-playing}~\cite{genesereth2013international}. For many domains hand-coding those models is an arduous task that faces {\em the knowledge acquisition bottleneck}, so inducing them from examples becomes an appealing approach. {\em Neural networks} (NNs) have proved to be effective learning transition models~\cite{schrittwieser2020mastering}. Unfortunately NNs represent knowledge as a huge number of coupled parameters, where it becomes challenging to identify the piece of knowledge responsible for modeling a particular feature of the examples, or to understand whether that knowledge will be useful at unseen examples.

In this paper we pursue a different research direction and focus on the synthesis of {\em white-box} models of state transitions, that are represented as {\em structured programs}. On the one hand programs of different kinds are proposed as an alternative to traditional action description languages~\cite{roger2008relative,katz2018semi,baier2022knowledge,Ronen:pprograms}. On the other hand, the flexibility of programming allows us to address model acquisition tasks with different {\em target languages}, without modifying the underlying method; methods for acquiring white-box models usually assume a particular {\em target language}, whose syntax and semantics is known, and that it is exploited to confine the hypothesis space of the possible models~\cite{geffner2022target}. A prominent example are systems for modeling {\em planning actions} that take \strips\ as their {\em target language}~\cite{yang2007learning,cresswell2013acquiring,kuvcera2018louga,aineto2019learning,asai2020learning,juba2021safe,rodriguez2021learning,lamanna2021online}, and that struggle when modeling systems whose state transitions do not fit the assumptions of the target language. 

\begin{figure}[t]
\includegraphics[scale=0.5]{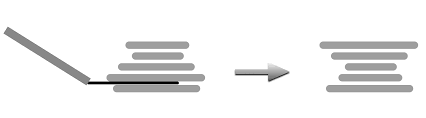}
\caption{Example of a state-transition in a five-pancake instance of the {\em pancake sorting} domain.}    
\label{fig:pancake:flip}
\end{figure}

We illustrate the previous issue in the  {\em pancake sorting} domain~\cite{cibulka2011average}, where a disordered stack of pancakes must be sorted, in order of size, using a spatula; the spatula can be inserted at any point in the stack to {\em flip} all the pancakes above. Figure~\ref{fig:pancake:flip} shows a state-transition of a {\em pancake sorting} instance. \strips\ model learners struggle at this domain because state transitions cannot be modeled by a single \strips\ action schema; \strips\ assumes that the number of state variables checked and updated at a state transition is bounded (and small). However in {\em pancake sorting}, the number of pancakes to flip is unbound, it depends on the number of world objects. Further, a compact procedural model requires {\em latent variables} that are missing in the state representation, since they do not describe observed properties/relations of the world objects. Figure~\ref{fig:pancake:flip:code} shows the {\tt flip} procedure, a compact model of the state transitions of the {\em pancake sorting} domain that leverages two {\em latent variables};  the {\tt flip} procedure models any state-transition, no matter the number of pancakes. 

The paper introduces a general approach for synthesising white-box models of the state transitions of a given discrete system, represented as structured programs. This synthesis problem is challenging, since the set of state transitions of a given discrete system may be infinite e.g. in domains like {\em pancake sorting } where the number of world objects is unbound.  Our approach is general since it accepts different {\em target languages}; relevant model acquisition tasks with different target languages, such as modeling \strips{} actions or the update rule of a {\em cellular automaton}~\cite{wulff1992learning}, fit as particular instances of our general approach.  Our synthesis method implements a combinatorial search in the space of well-structured terminating programs that can be built using a {\em Random-Access Machine} (RAM), with a minimalist instruction set, and a finite amount of memory. This method is able to synthesize compact white-box transition models, even for domains that require {\em latent variables}. In addition our method can exploit a {\em target language}, when available, to confine the solution space, without modifying the synthesis algorithm. 

\begin{figure}[t]
\centering
\scriptsize
\begin{lstlisting}[language=c++,numbers=none,mathescape]
State flip(State pre_state, int z1, int z2){
    State post_state=pre_state;
    for(z1=0; z1<$|\Omega|$; z1++){
        if (z1 < z2){
           post_state(z1) = pre_state(z2);
           post_state(z2) = pre_state(z1);
           z2 = z2 - 1;
        }
    }
    return post_state;
}
\end{lstlisting}
    \caption{Structured program that models state-transitions in the {\em pancake sorting} domain, no matter the number of pancakes $|\Omega|$, and that leverages two {\em latent variables} $\{z_1,z_2\}$. States are represented as tuples (whose length is the number of pancakes), that store the size of the pancake at the corresponding position in the stack.}   
\label{fig:pancake:flip:code}     
\end{figure}

\section{Preliminaries}

\subsection{Deterministic Transition Systems}
The notion of {\em transition system} is widely used in AI to model the behavior of discrete systems~\cite{baier2008principles}. A transition system can be graphically represented as a directed graph and hence formalized as a pair $(S,\rightarrow)$, where $S$ is a set of states, and $\rightarrow$ denotes a relation of state transitions $S\times S$. Transition systems differ from {\em finite automata} since the sets of states and transitions are not necessarily finite. Further a {\em transition system} does not necessary define a {\em start/initial} state or a subset of {\em final/goal} states.

Transitions between states may be labeled\footnote{When the set of labels is a singleton, the transition system is essentially {\em unlabeled}, so the simpler definition that omits  labels applies.}, and the same label may appear on more than one transition; a prominent example is the transition system that corresponds to a {\em classical planning problem}~\cite{ghallab2004automated,geffner2013concise}, where state transitions are labeled with {\em actions} s.t. between two different states $s, s'\in S$, there exists a transition $(s\xrightarrow{a}s')$ iff the execution of action $a$ in state $s$ produces the state $s'$.  Given a state $s$ and an action label $a$, if there exists only a single tuple $(s,a,s')$ in $\rightarrow$ then the transition is said to be {\em deterministic}. In this paper we focus on {\em deterministic} transitions systems, i.e. transition systems such that all their transitions are deterministic. 

\subsection{The Random Access Machine} 
The {\em Random-Access Machine} (RAM) is an abstract computation machine, in the class of the {\em register machines}, that is polynomially equivalent to a {\em Turing machine}~\cite{boolos2002computability}. The RAM enhances a multiple-register {\em counter machine}~\cite{minsky1961recursive} with indirect memory addressing; indirect memory addressing is useful for defining RAM programs that access an unbounded number of registers, no matter how many there are. A {\em register} in a RAM machine is then a memory location with both an {\em address} i.e. a unique identifier that works as a natural number (that we denote as $r$), and a {\em content} i.e. a single natural number (that we denote as $[r]$).

A {\em RAM program} $\Pi$ is a finite sequence of $n$ instructions, where each program instruction $\Pi[i]$, is associated with a {\em program line} {\small $0\leq i< n$}. The execution of a RAM program starts at its first program instruction $\Pi[0]$. The execution of program instruction $\Pi[i]$ updates the RAM {\em registers} and the {\em current program line}. Diverse {\em base instructions sets}, that are Turing complete, can be defined. We focus on the three {\em base sets} of RAM instructions:
\begin{itemize}
    \item Base1. $\{{\tt\small inc}(r)$, ${\tt\small dec}(r)$, ${\tt\small jmpz}(r,i)$, ${\tt\small halt}()$ $| \; r \in R\}$. Respectively, these instructions {\em increment}/{\em decrement} a register by one, jump to program line $0\leq i< n$ if the content of a register $r$ is zero (i.e. if $[r]== 0$), or end the program execution.
    \item Base2. $\{{\tt\small inc}(r_1)$, ${\tt\small clear}(r_1)$, ${\tt\small jmpz}(r_1,r_2,i)$, ${\tt\small halt}()$ $| \; r_1,r_2 \in R\}$. In this set the value of a register cannot be decremented but instead, it can be set to zero with a {\em clear} instruction. In addition, {\em jump instructions} go to program line $0\leq i< n$ if the content of two given registers is the same (i.e. if $[r_1]==[r_2]$).   
    \item Base3. $\{{\tt\small inc}(r_1)$, ${\tt\small set}(r_1,r_2)$, ${\tt\small jmpz}(r_1,r_2,i)$, ${\tt\small  halt}()$ $| \; r_1,r_2 \in R\}$. This set comprises no instruction to decrement, or clear, a register but instead, it includes an instruction to {\em set} a register to the value of another register.
    \end{itemize}
The three {\em base sets} are equivalent~\cite{boolos2002computability}; one can build the instructions of one base set with instructions of another base set. Further, {\em the expansive instruction set} (that contains the instructions of {\em Base} 1,2 and 3) does not increase the expressiveness of the individual {\em Base sets}, since each of them already is Turing complete. The choice of the set of RAM instructions depends on the convenience of the programmer for the problem being addressed.

\section{Transition Systems as RAM Programs}
This section formalizes our  representation for the state-transitions of a given deterministic discrete system.

\subsection{The RAM model} 
WLOG we assume that the states of a transition system are factored; given a set of world objects $\Omega$, a {\em state} is factored into a finite set of variables $X$ s.t. each variable $x\in X$ either represents a {\em property} of a world object, or a {\em relation} over $k$ world objects. Formally $x=f(o_1,\ldots,o_k)$, where $f$ is a $k$-ary function in $\mathbb{N}$, and $\{o_i\}_1^k$ are objects in $\Omega$. We compactly model the set of state {\em transitions} of a given discrete system as RAM programs $\Pi(s)=s'$ that map a given {\em pre-state} into its corresponding {\em post-state}. Next we formalize our particular RAM model for the compact representation of the state-transitions of a given discrete system.

Our RAM model partitions {\bf the set of RAM registers} into four disjoint subsets, $R=\{R_X,R_{X'},R_Y,R_Z\}$:
\begin{itemize}
    \item $R_X$ and $R_X'$, that are the {\bf pre-state} and {\bf post-state registers}. Our RAM model contains, for each state variable $x\in X$, two registers $r_x\in \mathcal{R}_X$ and $r_x'\in \mathcal{R}_{X'}$ representing the value of that variable at the {\em pre-state} and the {\em post-state}.   
    \item $R_Y=\{CF,ZF\}$, the {\bf FLAGS registers}. Our RAM model includes two dedicated Boolean registers, the {\em zero flag} (ZF) and the {\em carry flag} (CF),  storing the outcome of {\em three-way comparisons}~\cite{browning2020working} between two  registers. FLAGS registers allow us to keep the RAM instruction set compact,  reducing the set of {\em jump} instructions; the four joint values of $\{ZF,CF\}$ can model a large space of state features, including~$=\,$0, $\neq\,$0, $<0, >0, \leq 0, \geq 0$ as well as relations $=, \neq, <, >, \leq, \geq$ on pairs of registers. 
    \item $R_Z$, the {\bf latent registers} that play the role of auxiliary variables.   {\em Latent registers} allow to writing programs that implement sequences of state updates of unbounded size (a key property for computing compact transition models that apply no matter the actual number of world objects). 
\end{itemize}
Our RAM model has then $2|X|+2+|R_Z|$ registers, where $|X|=|R_X|=|R_X'|$ is the number of state variables. 

The {\bf instruction set} of our RAM model is $\{{\tt\small inc}(r)$, ${\tt\small dec}(r)$, ${\tt\small test}(r)$,  ${\tt\small set}(r,0|1)$, ${\tt\small set}(r_1,r_2)$, ${\tt\small cmp}(r_1,r_2)$,  ${\tt\small jmp(\neg ZF,\neg CF,i)}$, ${\tt\small jmp(\neg ZF,CF,i)}$, ${\tt\small jmp(ZF,\neg CF,i)}$, ${\tt\small jmp(ZF,CF,i)}$, ${\tt\small halt()}\}$. Respectively, these RAM instructions {\em increment}/{\em decrement} by one a register, {\em test} whether the value of a register is zero (or greater than zero), {\em set} the value of a register to {\tt zero}, {\tt one}, or to another register, {\em compare} two registers (or their content), jump to  program line $0\leq i< n$ according to the joint value of the FLAGS registers, or end the program. To keep the instruction set as compact as possible, and hence the space of candidate RAM programs, certain RAM instructions are only applicable to a subset of the RAM registers. In more detail, {\em pre-state registers} $R_X$  are {\em read-only}, so $test$ and $cmp$ instructions, are the only RAM instructions that apply to them. Likewise the {\em post-state registers} $R_X'$ are {\em write-only}, so test and comparisons cannot be defined over them. FLAGS registers are dedicated to store the outcome of {\em three-way comparisons}, so only jump instructions can be applied to them. The value of FLAGS registers is given by the outcome of RAM instructions; each RAM instruction  updates the FLAGS according to the result of the corresponding RAM instruction (which is denoted here by $res$):
\begin{scriptsize}    
\begin{align*}
 inc(r) &\implies res := r + 1,\\
 dec(r) &\implies res := r - 1,\\
 test(r) &\implies res := r,\\ 
 set(r_1,r_2) &\implies res := r_2,\\
 cmp(r_1,r_2) &\implies res := r_1 - r_2,\\
 ZF &:= ( res == 0 ),\\
 CF &:= ( res > 0 ).
\end{align*}
\end{scriptsize}

\subsection{The space of  RAM programs}
We consider RAM programs $\Pi: R\rightarrow R_{X'}$, that map a given {\em pre-state} into its corresponding {\em post-state}, with the assistance of the FLAGS and the {\em latent} registers. In this particular kind of RAM program, registers $R_{X'}$ are initialized with the value of their corresponding {\em pre-state} register, while FLAGS and {\em latent} registers are initialized to zero.  We restrict ourselves to $\Pi: R\rightarrow R_{X'}$ RAM programs that are both {\em well-structured} and {\em terminating}: 
\begin{itemize}
    \item {\bf Well-structured}. We only consider RAM programs whose jump instructions do not interleave, since this particular kind of program is more intelligible. Formally, given two program instructions $\Pi[i]=\mathsf{jump}(*,i')$ and $\Pi[j]=\mathsf{jump}(*,j')$ s.t.  $i<j$, it cannot hold that $min(i,i')<min(j,j')<max(i,i')<max(j,j')$. 
    \item {\bf Terminating}. WLOG we restrict ourselves to RAM programs that are by definition terminating. In more detail, RAM programs where any loop caused by a jump instruction is implementing a {\tt for loop} that iterates over the set of world objects. Formally, we only consider RAM programs where any {\em jump instruction} that  corresponds to a loop (i.e. such that $\Pi[i]=\mathsf{jump}(*, i')$ and $i'<i$) iterates over the  set of world-objects i.e. loops of the form $\Pi[i]=set(r,0)$, $\Pi[i+k]=inc(r)$, $\Pi[i+k+1]=jmpz(ZF,CF,i)$. 
\end{itemize}

Restricting to {\em well-structured} and {\em terminating} RAM programs, allows us to keep the space of candidate RAM programs compact. It also improves the intelligibility of candidate RAM programs, since it allows to replace $test$, $cmp$ and $jmp$ instructions by their equivalent {\tt If conditionals} and {\tt For loops} control flow constructs from {\em structured programming}; in a {\em well-structured} and {\em terminating} RAM program a $\Pi[i]=\mathsf{jump}(*,i')$ instruction always  represents either an {\em if conditional}   (when $i'>i$) or a {\em for loop}  (when $i'<i$). 

Now we are ready to formulate the formal grammar that defines the  space of {\em well-structured} and {\em terminating} RAM programs. In this grammar the non-terminal symbol {\em RAMInstruction} refers to an instruction from our set of primitive RAM instructions (in more detail, refers to an {\em inc}, {\em dec} or {\em set} instruction since $test$, $cmp$ and $jmp$ instructions are  replaced by their corresponding control flow constructs from {\em structured programming}). Please note that the grammar accepts the structured program of Figure~\ref{fig:pancake:flip:code}; we exemplify our representation with a generic high-level structured programming language-like, that supports {\tt If} conditionals and {\tt For} loops, as well as {\tt tuples} (to store the state variables) as could be the case of common structured programming languages like {\em Python}, {\em C++} or {\em Java}. 

\begin{scriptsize}
\begin{align*}
\Pi \Coloneqq\  & post\_state=pre\_state;\\
& Instruction(s)\\
& return\ post\_state;\\
Instruction(s) \Coloneqq\ & If; Instruction(s)\mid\\ 
& For; Instruction(s)  \mid\\
& RAMInstruction ; Instruction(s)\mid\\ 
& ;\\
If \Coloneqq \ & if(Condition)\{Instruction(s)\}\\
Condition \Coloneqq \ & (r>0)\mid (r==0)\mid (r_1>r_2)\mid (r_1==r_2)\mid (r_1<r_2)\\ 
For \Coloneqq\ & for(r=0;r<|\Omega|;r++)\{Instruction(s)\}|\\
& for(r=|\Omega|-1;z\geq 0;r--)\{Instruction(s)\}\\
\end{align*}
\label{eq:GPgrammar}
\end{scriptsize}

{\bf Example.} Our RAM model for the {\em pancake sorting} domain has $2\times|\Omega|+2+2$ registers, where $|R_X|=|R_X'|=|\Omega|$ indicates the number of pancakes. State variables are represented with $|\Omega|$ RAM registers $r=f_i(o)$, with domain $D_r=[0,\Omega)$, where $f_i(o)$ indicates the size of  pancake $o\in \Omega$ located the $i^{th}$ at the stack of pancakes.  In this domain only $1-$ary properties of the pancakes are necessary to represent states, so one can use a single auxiliary RAM register to enumerate the pancakes and iterate over the state variables; Figure~\ref{fig:pancake:flip:code}, showed a {\em well-structured} and {\em terminating} RAM program that enumerates the pancakes with  the {\em latent register} $z_1\in R_Z$, no matter the actual number of pancakes. 

\section{Target Languages as RAM Program Spaces}
When available, a particular {\em target language} can be exploited to implement more effective enumerations of the space of candidate RAM programs. This section shows that different formal grammars can be defined over our RAM {\em instruction set} to exploit the particular structure of relevant modeling tasks, such as modeling \strips\ planning actions or the update rule of a {\em cellular automaton}.

\subsection{Target Language 1: \strips{}}
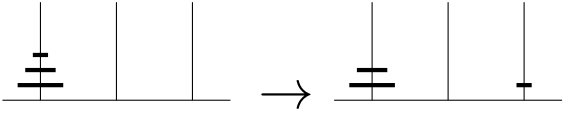
\begin{figure}[t]
\begin{tikzpicture}[block/.style= {rectangle, draw=black, thick, text centered, node distance=.4cm}]
\draw [ultra thick] (-0.1,-0.4) -- (0.1,-0.4) node [] {};
\draw [ultra thick] (-0.2,-0.6) -- (0.2,-0.6) node [] {};
\draw [ultra thick] (-0.3,-0.8) -- (0.3,-0.8) node [] {};
\draw [thin] (-.5,-1) -- (2.5,-1) node [] {};
\draw [thin] (0,-1) -- (0,0.3) node [] {};
\draw [thin] (1,-1) -- (1,0.3) node [] {};
\draw [thin] (2,-1) -- (2,0.3) node [] {};
\end{tikzpicture}
{\Huge \ $\rightarrow$\ }
\begin{tikzpicture}[block/.style= {rectangle, draw=black, thick, text centered, node distance=.4cm}]
\draw [ultra thick] (-0.2,-0.6) -- (0.2,-0.6) node [] {};
\draw [ultra thick] (-0.3,-0.8) -- (0.3,-0.8) node [] {};
\draw [ultra thick] (1.9,-0.8) -- (2.1,-0.8) node [] {};

\draw [thin] (-.5,-1) -- (2.5,-1) node [] {};
\draw [thin] (0,-1) -- (0,0.3) node [] {};
\draw [thin] (1,-1) -- (1,0.3) node [] {};
\draw [thin] (2,-1) -- (2,0.3) node [] {};
\end{tikzpicture}
  \caption{Example of a state-transition in a three-disk instance of the {\em Tower of Hanoi} domain, where the smallest disk is moved from the leftmost peg to the rightmost peg.} 
  \label{fig:tower}
\end{figure}

{\em Classical planning} addresses the computation of sequences of deterministic actions that transform an initial state, into a state that satisfies a given set of goals. Figure~\ref{fig:tower} illustrates a state-transition from a three-disk classical planning instance of the {\em Tower of Hanoi} (ToH). The infinite set of state transitions of the classical planning domain of the ToH are compactly modeled with a single \strips\ operator (Figure~\ref{fig:tower:move:strips}), no matter the actual number of disks. The {\tt move} operator leverages three first-order predicates that indicate: whether an object is {\tt on} top of another object, whether an object is {\tt smaller} than another object and whether an object is {\tt clear} i.e. has no other object on top of it.

Let $\Psi$ be the set of first-order k-arity predicates of a classical planning domain, then $\Psi(X)$ is the set of atomic formulas $f(x_1,\ldots, x_k)$ obtained by applying the k-ary predicates $f\in \Psi$ to any m-tuple $X=\tup{x_1,\ldots, x_m}$  of distinct symbols (either constants or variables), with $1\leq \ldots, m\leq k$. For instance, if $\Psi$ contains the single binary predicate {\tt on}, and $X=\tup{x_1, x_2, x_3}$. Then, $\Psi(X) = \{on(x_i, x_j ) | 1 \leq i, j \leq 3\}$. A k-ary \strips\ operator $op\in \mathcal{O}$, defined on the set of predicates $\Psi$, is a tuple $\tup{par(op), pre(op), eff^+(op), eff^-(op)}$, where $par(op)$ is a k-tuple of variables, $pre(op)$, $eff^+(op)$, and $eff^-(op)$ are three sets of atoms on $\Psi(par(op))$ representing the {\em preconditions}, {\em positive}, and {\em negative effects} of the operator. WLOG, the \strips\ formalism assumes that effects are consistently defined, i.e. $eff^-(op) \subseteq pre(op)$ and $eff^+(op) \bigcap eff^-(op) = \emptyset$. 

\begin{figure}[t]
\centering
\scriptsize
\begin{lstlisting}[language=lisp,numbers=none]
(:action move 
 :parameters (?disc ?from ?to)
 :precondition (and (smaller ?to ?disc) (on ?disc ?from) 
                    (clear ?disc) (clear ?to))
 :effect  (and (clear ?from) (on ?disc ?to) 
               (not (on ?disc ?from)) (not (clear ?to)))
\end{lstlisting}
    \caption{\strips\ operator that models all the state transitions of the {\em Tower of Hanoi}, no matter the number of disks.}   
\label{fig:tower:move:strips}     
\end{figure}  

Next, we formulate our RAM representation for the state transitions of a \strips\ system:
\begin{itemize}
    \item {\bf States}. Each ground predicate $p(o_1,\ldots, o_k)$ is represented by one RAM register $r\in R_X$, with domain $D_r=\{0,1\}$, and another one in $R_X'$. This state representation requires then a tuple of $|R_X|=|R_X'|=\sum_{k\geq 0} n_k|\Omega|^k$ Boolean registers, where $n_k$ is the number of first-order predicates with arity $k$, and $|\Omega|$ is the number of world objects\footnote{ $\sum_{k\geq 0} n_k|\Omega|^k$ is also the  number of propositions that result from grounding a \strips\ classical planning instance.}. The RAM model defines also $|R_Z|={\sf max}_{p\in\Psi} arity(p)$ latent registers, i.e. given by the largest arity of a predicate in $\Psi$. The {\em latent registers} have domain $D_z=[0,\Omega)$ so they can index registers in $R_X$ and $R_X'$. 
    \item {\bf Transitions}. Each \strips\ operator is represented as a $\Pi: R\rightarrow R_{X'}$ RAM program that first, it copies the {\em pre-state} into the {\em post-state}. Then the program checks whether the preconditions of the \strips\ operator hold in the {\em pre-state}, and iff this is the case, it updates and returns the {\em post-state}. Figure~\ref{fig:hanoi:move} shows a RAM program modeling the {\tt move} \strips\  operator, that represents the state-transitions of the ToH, no matter the number of disks. 
\end{itemize}

Next we show the formal grammar that confines the space of candidate RAM programs to the particular structure of \strips.

\begin{scriptsize}
\begin{align*}
\Pi \Coloneqq\  & post\_state=pre\_state;\\ 
& Instruction(s)\\ 
& return\ post\_state;\\
Instruction(s) \Coloneqq\ & If; Instruction(s);\\
If \Coloneqq \ & if(Condition)\{Instruction(s)\}\mid\\
& if(Condition)\{Assigment\}\\
Condition \Coloneqq \ & (pre\_state(f,z_1,\ldots, z_k)==0)\mid\\
& (pre\_state(f,z_1,\ldots, z_k)==1)\\
Assigment(s) \Coloneqq \ & post\_state(f,z_1,\ldots, z_k)=0;\ Assigment(s)\mid\\
& post\_state(f,z_1,\ldots, z_k)=1;\ Assigment(s)\mid\\
&;
\end{align*}
\end{scriptsize}

\noindent where $Condition(s)$ is a conjunction of assertions over predicates $p(z_1,\ldots, z_k)$ instantiated with the action arguments (i.e. the latent registers $z\in R_Z$), and representing the operator preconditions ($==$ denotes the equality operator, $=$ indicates an assignment, and a semicolon denotes the end of a program instruction). {\em Assignment(s)} is a conjunction of assignments  representing the operator positive/negative effects;  in more detail $p(z_1,\ldots, z_k) = 1$ denotes a {\em positive} effect while $p(z_1,\ldots, z_k) = 0$
denotes a {\em negative} effect.

\begin{figure}[t]
\centering
\scriptsize
\begin{lstlisting}[language=c++,numbers=none,mathescape]
State move(State pre_state, int disc, int from, int to){
   State post_state=pre_state;
   if (pre_state(smaller,to,disc) == 1){
       if (pre_state(on,disc,from) == 1){
           if (pre_state(clear,disc) == 1){
               if (pre_state(clear,to) == 1){
                   post_state(clear,from)=1;
                   post_state(on,disc,to)=1;
                   post_state(on,disc,from)=0;
                   post_state(clear,to)=0;
    }}}}
    return post_state;
}
\end{lstlisting}
    \caption{RAM program modeling the {\tt\small move} operator from the {\em Tower of Hanoi}.} 
\label{fig:hanoi:move}     
\end{figure}

{\bf Universally Quantified Conditional Effects. } \strips\ biases modeling towards state transitions that only require checking and updating a fixed (and small) number of state variables. As a matter of fact, this number is actually upper bounded by the {\em header} (number and type of parameters) of the corresponding \strips\ operator. Our approach goes beyond \strips\ and can also naturally model actions with {\em universally quantified preconditions} and {\em conditional effects} that are able to check and update and unbound number of state variables; that is the particular case of the fragment of the PDDL planning language~\cite{haslum2019introduction} that is obtained by setting the {\small\tt :universal-preconditions} and {\tt\small :conditional-effects} requirements. 
In this case our state representation is the same as for STRIPS systems but the target language is extended with {\em for loop} instructions from RAM programs, that leverage latent variables to iterate over the set of world objects.

\subsection{Target language 2: Cellular automata}
A cellular automaton is a {\em zero-player game}, defined as a collection of finite-domain variables, called {\em cells}, that are situated on a grid of a defined shape; the grid can be defined in any finite number of dimensions and the number of cells is unbounded. The initial state of a cellular automaton is set  assigning a value to every cell in the automaton. The state transitions of a cellular automaton are given by a fixed set of {\em update rules}, that define the next value of a cell w.r.t.: (i), its current value and (ii), the current value of its neighbour cells. The simplest non-trivial cellular automaton is {\em one-dimensional}, with two possible values per cell $\{0,1\}$, and the cell's neighborhood defined as the  adjacent cells on either side of a given cell. A cell and its two neighbors form then a three-cell neighborhood, so there are $2^3 = 8$ possible {\em neighborhood patterns}, and  $2^8 = 256$ different one-dimensional cellular automata, generally named by their {\em Wolfram code}~\cite{wolfram2002new}. Table ~\ref{tab:rule30} shows the {\em Rule 30} cellular automaton.

\begin{table}
\scriptsize
\begin{tabular}{l|c|c|c|c|c|c|c|c}
Neighbrhd. pattern & 111 & 110 & 101 & 100 & 011 & 010 & 001 & 000 \\\hline
Next value & 0 & 0 & 0 & 1 & 1 & 1 & 1 & 0
\end{tabular}
\caption{{\em Rule 30} cellular automaton. For each neighborhood pattern, the rule specifies the next value for the center cell.}
\label{tab:rule30}
\end{table}

\begin{figure}[t]
\centering
\scriptsize
\begin{lstlisting}[language=c++,numbers=none,mathescape]
State R30(State pre_state, int z1, int z2, int z3){
    State post_state=pre_state;
    for(z1=0; z1<$|\Omega|$; z1++){
        z2=z1; z2--;
        z3=z1; z3++;
        if(pre_state(z1)==1){
            if(pre_state(z2)==1){
                if(pre_state(z3)==1){
                    post_state(z1)=0;
                }}}
                
        ...
        
        if(pre_state(z1)==0){
            if(pre_state(z2)==0){
                if(pre_state(z3)==0){
                    post_state(z1)=0;
                }}}
    }
    return post_state;
}
\end{lstlisting}
    \caption{Fragment of RAM program modeling the {\em Rule 30} automaton that leverages three {\em latent registers} $R_Z=\{z_1,z_2,z_3\}$.}   
\label{fig:r30:code}     
\end{figure}

We formulate our RAM representation for the state transitions of a one-dimensional cellular automaton as follows:
\begin{itemize}
    \item {\bf States}. The RAM representation for the state of a one-dimensional cellular automaton is a tuple of $|R_X|=|R_X'|=|\Omega|$ Boolean registers (i.e. one register per each cell in the automaton), plus three {\em latent registers} $R_Z=\{z_1,z_2,z_3\}$, with domain $D_z=[0,\Omega)$, to index the registers corresponding to a given cell and its two  neighbour cells. 
    \item {\bf Transitions}. The set of update rules of a one-dimensional cellular automaton is represented as a $\Pi: R\rightarrow R_{X'}$ RAM program that first, it copies the {\em pre-state} into the {\em post-state}. Then, for each cell in the automaton, the program updates the cell value according to the value of its corresponding neighborhood. Figure~\ref{fig:r30:code} shows a fragment of the RAM program that models the state-transitions of the {\em Rule 30} automaton (no matter its number of cells). 
\end{itemize}

Next, we show the formal grammar that confines the space of RAM programs to the particular structure of {\em one-dimensional cellular automata}. 

\begin{tiny}
\begin{align*}
\Pi \Coloneqq\  & post\_state=pre\_state;\\ 
& for(z1=0;z1<|\Omega|;z1++)\{\\
& z2=z1; dec(z2); \\
& z3=z1; inc(z3); \\
& Instruction(s)\\
& \}\\
& return(post\_state);\\
Instruction(s) \Coloneqq\ & If; Instruction(s);\\
If \Coloneqq \ & if(Condition)\{Instruction(s)\}\mid\\
& if(Condition)\{Assigment\}\\
Condition \Coloneqq \ & (pre\_state(z)==0)\mid (pre\_state(z)==1)\\
Assigment \Coloneqq \ & post\_state(z1)=0;\mid\\
& post\_state(z1)=1;
\end{align*}
\label{eq:GPgrammar:CA}
\end{tiny}

\section{Synthesis of Procedural Models}
This section formalizes the problem of synthesizing  structured programs that model the {\em state-transitions} of a given {\em deterministic transition system}. Then, the section details our synthesis method and its theoretical properties.

\subsection{The problem}
The {\em problem} of synthesizing  the RAM programs that model the {\em state-transitions} of a  {\em deterministic transition system} is formalized as a tuple $\tup{\mathcal{E},\mathcal{M}}$, where:
\begin{itemize}
    \item $\mathcal{E}$ is a {\em set of  examples}, s.t. each example $e=(s_t,a,s_{t+1})$ comprises (i), a {\em pre-state} (ii), a {\em transition label} and (iii), its  {\em post-state}. We denote $\mathcal{E}_a \subseteq\mathcal{E}$, the subset of examples with transition label $a$.    
    \item $\mathcal{M}$ is a {\em RAM model} (as defined above) that comprises registers $R=\{R_X,R_{X'},R_Y,R_Z\}$, and s.t.  registers in $R_X$ (and in $R_{X'}$ too) model the state space of the {\em deterministic transition system}. 
\end{itemize}

A {\em solution} to the synthesis problem is a finite and non-empty set of RAM programs $\Pi_a: R\rightarrow R_{X'}$ s.t. for all the examples $e=(s_t,a,s_{t+1})$ in the subset $\mathcal{E}_a \subseteq\mathcal{E}$, the execution of $\Pi_a$ in the {\em pre-state} $s_t$, exactly outputs its corresponding {\em post-state} $s_{t+1}$. 

{\bf Example.} We exemplify the synthesis problem in the {\em pancake sorting} domain, with a set of three  examples $\mathcal{E}=\{e_1,e_2,e_3\}$ that  represent three transitions of a four-pancake instance; $e_1=((3,2,1,4),{\tt\small flip}(2),(1,2,3,4))$, $e_2=((2,3,1,4),{\tt\small flip}(1),(3,2,1,4))$ and $e_3=((1,3,2,4),{\tt\small flip}(2),(2,3,1,4))$. The RAM model $\mathcal{M}$ comprises four registers $R_X=\{r_{x_1},r_{x_2},r_{x_3},r_{x_4}\}$  encoding the {\em pre-state} of a transition (each register stores the size of the pancake at its corresponding position in the stack), four registers $R_{X}'=\{r_{x_1}',r_{x_2}',r_{x_3}',r_{x_4}'\}$  encoding the {\em post-state} of a transition, the two FLAGS registers $R_Y=\{CF,ZF\}$, two latent registers $R_Z=\{z_1,z_2\}$, and the RAM instruction set defined above.

\subsection{The  method}
Our synthesis  method implements a {\em Best First Search} (BFS) in the space of  {\em well-structured} and {\em terminating} RAM programs, that can be built with a maximum number of program lines $n$ and latent registers $|R_z|$. In more detail, each search node corresponds to a {\em partially specified} RAM program $\Pi_a: R\rightarrow R_{X'}$; by partially specified we mean that some of the program lines may be undefined, because they are not programmed yet. Starting from the {\em empty program} (i.e. the partially specified program whose all lines are undefined), the space of programs is enumerated with a search operator that programs an  instruction, at an undefined program line $0\leq i< n$. This search operator is only applicable when program line $i$ is undefined; initially $i:=0$, and after line $i$ is programmed $i:=i+1$. 

To reduce memory requirements we implement a {\em frontier BFS}~\cite{korf2005frontier}, that store only the open list of generated nodes but not the closed list of expanded nodes. Nodes in the open list are sorted w.r.t. the following evaluation functions:
        \begin{itemize}
            \item $f_{\#ifs}(\Pi_a)$, returns the number of {\em conditional if} instructions of a given RAM program $\Pi_a$.       
            \item $f_{\#loops}(\Pi_a)$, returns the number of {\em for loop} instructions of a given RAM program $\Pi_a$.
            \item $f_{gc}(\Pi_a,\mathcal{E}_a) = \sum_{e\in\mathcal{E}_a}\sum_{x\in s_{t+1}} |s_{t+1}[x]-s_{t+1}'[x]|$, where $\Pi_a$ is a  RAM program, $\mathcal{E}_a \subseteq\mathcal{E}$ is the subset of examples with label $a$, and $s_{t+1}'$ is the state produced by executing  $\Pi_a$ on the pre-state $s_t$ of a example $e\in\mathcal{E}_a$. For all the examples $e\in\mathcal{E}_a$, the function $f_{gc}$ accumulates the number of state variables in $s_{t+1}'$ whose value mismatch with the corresponding post-state $s_{t+1}$.
        \end{itemize}
        
In more detail, nodes in the open list are sorted maximizing the function $f_{\#loops}(\Pi_a)$, then breaking ties by maximizing $f_{\#ifs}(\Pi_a)$ and last, breaking ties by minimizing $f_{gc}(\Pi_a,\mathcal{E}_a)$. This combination of the evaluation functions biases synthesis towards models with larger number of conditions that cover the given set of  examples.  In addition we implement a {\em pruning rule} $R1$ that reduces the search space while it preserves the solution space (i.e. preserves completeness). The rule leverages the fact that between the {\em pre-state} and the {\em post-state} of an example there is no  intermediate states:
\begin{itemize}
\begin{small}
    \item $R1$. We do not allow programming RAM instructions that set a register in $R_X'$ with a value different from its value at the post-state of a given example.
\end{small}    
\end{itemize}

{\bf Properties.} Our method for synthesizing  RAM programs that model the set of {\em state-transitions} of a given {\em discrete system}  is terminating; termination follows from a terminating {\em search algorithm} and {\em evaluation functions}. Regarding the former, our method implements a {\em frontier BFS} which is known to be terminating at finite search spaces~\cite{korf2005frontier}, we recall that our search space is finite since the maximum number of program lines $n$ is bounded. Regarding the evaluation functions, $f_{\#loops}$ and $f_{\#ifs}$, they terminate in $n$ steps, where $n$ is the number of lines of the program to evaluate. Last, $f_{gc}$ terminates iff the program executions terminate, which is always true since our candidate RAM programs are by definition {\em well-structured} and {\em terminating}. Our method is {\em sound}, since it only outputs  RAM programs $\Pi_a$ that cover the full set of  examples $\mathcal{E}_a$. Last, our method is {\em complete} provided that there exists a solution within the given number of program lines $n$ and latent registers $|R_Z|$.

\section{Evaluation}
We report results on the {\bf synthesis} and {\bf validation} of procedural models for a wide range of {\em discrete systems}, and using the following {\em target languages}:
\begin{enumerate}
\item \strips. We synthesize 49 action schemes from 13 domains ({\em FD benchmarks}\footnote{https://github.com/aibasel/downward-benchmarks} and {\sc Planning.Domains} \cite{muise2016planning}, ranging from {\em Gripper} to {\em Transport} in Table~\ref{tab:strips_actions}). For each domain, examples are generated using 10 IPC instances and computing: (i), one random walk per instance of at most 50 steps with {\sc Tarski} \cite{tarski:github:18} and (ii), solution plans for those instances with the {\sc LAMA}~\cite{richter2011lama} setting of {\sc FD}~\cite{helmert2006fast}. {\em Synthesis examples} ${\mathcal E}_{synth}$, and {\em validation examples} ${\mathcal E}_{test}$, are the traversed {\em (pre-state, action, post-state)} tuples. 
\item \strips{} {\em with universally quantified conditional effects}. We synthesize  7 action schemes from 3 different classical planning domains (namely, {\em briefcase}, {\em maintenance}, and {\em elevators}) that contain actions with universally quantified conditional effects. Examples are generated as for the \strips{} domains. 
\item {\em Cellular automaton}. We synthesize the update rule of four well-known one-dimensional {\em cellular automaton}~\cite{wolfram2002new}, namely rule {\em 30}, rule {\em 90}, rule {\em 110} and rule {\em 184}. For each rule,  ${\mathcal{E}_{synth}}$ examples are 20 state-transitions of a 19-cell automaton which starts with 3 ones in the three central cells, while  ${\mathcal{E}_{test}}$ examples are 100 state-transitions of 99-cell automata. 
\item {\em RAM}. We synthesize the program that models the {\small\tt flip} action scheme from {\em pancake sorting}; ${\mathcal{E}_{synth}}$ examples comprise 16 flips of a randomly generated 9-pancake instance, while ${\mathcal{E}_{test}}$ examples comprise 98 flips of a randomly generated 50-pancake instance.
\end{enumerate}
For \strips, \strips{} {\em with quantified effects} and the {\em Cellular automaton} domains, our synthesis method exploits their corresponding grammars that confine the space of possible  programs. For the {\em pancake sorting} we consider the full space of RAM programs. All experiments  are performed in an Ubuntu 22.04.2 LTS, with Intel® Core i5-10300H @ 2.50GHz x 8-core processor and 16GB of RAM\footnote{Framework: https://github.com/jsego/bfgp-pp}. 

\textbf{Synthesis.} Our synthesis method is fed with two input parameters: (i) the maximum number of {\em program lines} $n$ and (ii), the maximum number of {\em latent variables} $|R_Z|$. Target languages like \strips{} or \textit{Cellular Automaton} have a known upper-bound in the number of required program lines. In these domains we used those upper-bounds otherwise (in {\em pancake sorting} and domains with quantified effects) we incrementally increase program size until a program is synthesized that satisfies the input examples; similar to what is done in SAT-planning with the {\em planning horizon}~\cite{kautz1999unifying}. The first six columns of Table~\ref{tab:strips_actions} report the obtained {\em synthesis results}. For each domain and action schema we report, the number of synthesis examples $|\mathcal{E}_{synth}|$, the number of lines in the solution ($n_{sol}$) out of upper-bound $n_{max}$, the time elapsed $T_{synth}$, and the number of expanded (Exp.) and evaluated (Eval.) search nodes.  Note that $T_{synth}$ takes less than a sec in almost half of the cases, and less than a minute in all of them, even when $|{\mathcal E}_{synth}|$ comprises several hundreds of examples; we effectively deal with hundreds of synthesis examples progressively adding input examples to the synthesis set, so they may act as counter-examples of candidate solutions~\cite{segovia2022scaling}. Results show that in \strips{} domains the node to expand is perfectly selected by BFS, since $n_{sol}$ equals the number of expansions, only loosing time in node evaluation. For the rest of languages, there are  3 cases where expansions go beyond 10K nodes, but still keeping a good trade-off between time and memory (expanded and evaluated are nodes are close).

\begin{table*}[t]
    \scriptsize
    	\centering
    		\begin{tabular}{|l|l||c|c|c|c||c|c|c|} \hline
          		{\bf Domain} & {\bf Action Schema} & $|\mathcal{E}_{synth}|$ & $n_{sol}/n_{max}$ & $T_{synth}$ & Exp./Eval. & 
          		$|\mathcal{E}_{test}|$ &  $T_{test}$& $\%\checkmark$\\\hline
\multirow{ 3 }{*}{ gripper } & move & 138 & 31/82 & 0.45 & 31/554 & 231 & 0.12& $100\%$\\ & pick & 116 & 58/82 & 1.21 & 58/1271 & 203 & 0.11& $100\%$\\ & drop & 111 & 58/82 & 1.11 & 58/1271 & 201 & 0.11 & $100\%$\\\hline
\multirow{ 4 }{*}{ miconic } & depart & 40 & 35/49 & 1.54 & 35/458 & 66 & 0.21& $100\%$\\ & up & 162 & 35/49 & 5.09 & 35/458 & 172 & 0.55 & $100\%$\\ & down & 137 & 35/49 & 4.37 & 35/458 & 168 & 0.52& $100\%$\\ & board & 64 & 30/49 & 1.74 & 30/359 & 107 & 0.33& $100\%$\\\hline
\multirow{ 6 }{*}{ driverlog } & disembark-truck & 29 & 94/241 & 3.58 & 94/4977 & 46 & 0.07& $100\%$\\ & board-truck & 34 & 94/241 & 3.90 & 94/4977 & 54 & 0.10& $100\%$\\ & drive-truck & 42 & 163/241 & 10.89 & 163/9962 & 80 & 0.13& $100\%$\\ & walk & 90 & 93/241 & 6.16 & 93/4944 & 88 & 0.14& $100\%$\\ & unload-truck & 67 & 91/241 & 4.94 & 91/4724 & 71 & 0.10& $100\%$\\ & load-truck & 73 & 91/241 & 5.24 & 91/4724 & 76 & 0.11& $100\%$\\\hline
\multirow{ 4 }{*}{ parking } & move-car-to-curb & 124 & 24/61 & 0.31 & 24/361 & 150 & 0.06& $100\%$\\ & move-curb-to-curb & 43 & 17/61 & 0.10 & 17/233 & 40 & 0.01& $100\%$\\ & move-curb-to-car & 135 & 22/61 & 0.37 & 24/410 & 153 & 0.06& $100\%$\\ & move-car-to-car & 144 & 29/61 & 0.52 & 29/440 & 158 & 0.12& $100\%$\\\hline
\multirow{ 1 }{*}{ visitall } & move & 548 & 14/19 & 5.37 & 14/79 & 1183 & 7.59& $100\%$\\\hline
\multirow{ 5 }{*}{ grid } & move & 187 & 47/232 & 26.14 & 47/2092 & 254 & 1.08& $100\%$\\ & unlock & 9 & 155/232 & 31.66 & 155/9085 & 8 & 0.05& $100\%$\\ & pickup & 33 & 50/232 & 8.82 & 50/2295 & 22 & 0.11& $100\%$\\ & putdown & 31 & 50/232 & 8.18 & 50/2295 & 21 & 0.09& $100\%$\\ & pickup-and-loose & 11 & 97/232 & 16.66 & 97/5086 & 15 & 0.08& $100\%$\\\hline
\multirow{ 4 }{*}{ blocks } & stack & 89 & 22/28 & 0.07 & 22/194 & 155 & 0.04& $100\%$\\ & unstack & 91 & 22/28 & 0.07 & 22/194 & 160 & 0.04& $100\%$\\ & put-down & 79 & 13/28 & 0.03 & 13/95 & 131 & 0.03& $100\%$\\ & pick-up & 77 & 13/28 & 0.03 & 13/95 & 126 & 0.03& $100\%$\\\hline
\multirow{ 5 }{*}{ satellite } & calibrate & 11 & 104/265 & 2.12 & 104/5918 & 9 & 0.03& $100\%$\\ & switch-on & 73 & 50/265 & 1.73 & 50/2546 & 91 & 0.14& $100\%$\\ & switch-off & 65 & 49/265 & 1.36 & 49/2459 & 68 & 0.11& $100\%$\\ & turn-to & 143 & 103/265 & 9.01 & 103/5915 & 520 & 0.93& $100\%$\\ & take-image & 32 & 174/265 & 9.01 & 174/11483 & 254 & 0.46& $100\%$\\\hline
\multirow{ 1 }{*}{ npuzzle } & move & 729 & 17/19 & 1.63 & 17/107 & 3390 & 5.17& $100\%$\\\hline
\multirow{ 1 }{*}{ hanoi } & move & 307 & 27/46 & 0.75 & 27/333 & 2229 & 1.57& $100\%$\\\hline
\multirow{ 9 }{*}{ rovers } & communicate-rock-data & 90 & 24/160 & 1.45 & 24/673 & 46 & 0.09& $100\%$\\ & calibrate & 92 & 14/160 & 0.56 & 14/353 & 61 & 0.12& $100\%$\\ & communicate-soil-data & 74 & 26/160 & 1.40 & 26/752 & 43 & 0.09& $100\%$\\ & sample-rock & 42 & 23/160 & 0.68 & 23/728 & 24 & 0.12& $100\%$\\ & communicate-image-data & 91 & 24/160 & 1.47 & 24/673 & 44 & 0.09& $100\%$\\ & sample-soil & 36 & 23/160 & 0.66 & 23/728 & 25 & 0.05& $100\%$\\ & navigate & 174 & 17/160 & 1.34 & 17/483 & 113 & 0.22& $100\%$\\ & drop & 51 & 11/160 & 0.29 & 11/289 & 32 & 0.07& $100\%$\\ & take-image & 70 & 17/160 & 0.62 & 17/475 & 44 & 0.09& $100\%$\\\hline
\multirow{ 3 }{*}{ ferry } & debark & 81 & 30/40 & 1.87 & 30/340 & 93 & 0.15& $100\%$\\ & sail & 216 & 27/40 & 3.54 & 27/285 & 239 & 0.39& $100\%$\\ & board & 81 & 30/40 & 1.77 & 30/340 & 96 & 0.16& $100\%$\\\hline
\multirow{ 3 }{*}{ transport } & drive & 217 & 11/40 & 0.40 & 11/89 & 427 & 0.58& $100\%$\\ & pick-up & 79 & 19/40 & 0.26 & 19/205 & 47 & 0.06& $100\%$\\ & drop & 76 & 19/40 & 0.25 & 19/205 & 45 & 0.07& $100\%$\\\hline\hline
\multirow{ 3 }{*}{ briefcase } & move & 167 & 15/15 & 2.27 & 27215/27245 & 278 & 0.07& $100\%$\\ & take-out & 68 & 4/4 & 0.01 & 3/4 & 119 & 0.03& $100\%$\\ & put-in & 71 & 8/8 & 0.01 & 7/18 & 124 & 0.03& $100\%$\\\hline
\multirow{ 3 }{*}{ elevators } & stop & 127 & 18/18 & 9.22 & 67283/67326 & 142 & 0.18& $100\%$\\ & up & 94 & 7/7 & 0.04 & 17/28 & 134 & 0.19& $100\%$\\ & down & 85 & 7/7 & 0.04 & 17/30 & 119 & 0.16& $100\%$\\\hline
\multirow{ 1 }{*}{ maintenance } & workat & 47 & 9/9 & 0.08 & 59/65 & 53 & 0.03& $100\%$\\\hline\hline
\multirow{ 4 }{*}{ cellular } & rule184 & 20 & 95/95 & 0.15 & 242/257 & 100 & 0.26& $100\%$\\ & rule30 & 20 & 95/95 & 0.09 & 155/257 & 100 & 0.25& $100\%$\\ & rule90 & 20 & 95/95 & 0.08 & 158/256 & 100 & 0.23& $100\%$\\ & rule110 & 20 & 95/95 & 0.11 & 189/257 & 100 & 0.22& $100\%$\\\hline\hline
\multirow{ 1 }{*}{ pancakes } & pancakes-flip & 16 & 8/8 & 12.17 & 20163/20837 & 98 & 0.07& $100\%$	
    		\\\hline\end{tabular}
    	\caption{For each domain and action schema we report, the number of synthesis examples $|\mathcal{E}_{synth}|$, the number of lines in the solution ($n_{sol}$) out of an upper-bound $n_{max}$, the time elapsed in seconds $T_{synth}$, the number of expanded (Exp.) and evaluated (Eval.) nodes by the BFS, the number of validation examples $|\mathcal{E}_{test}|$, the validation time in secs $T_{test}$ and the rate of $\mathcal{E}_{test}$ examples successfully validated $\%\checkmark$.}
    	\label{tab:strips_actions}
	\end{table*}

\textbf{Validation.} Validation examples in ${\mathcal E}_{test}$ are larger than the ${\mathcal E}_{synth}$ examples (i.e. larger number of objects). The three last columns of Table~\ref{tab:strips_actions} (namely, the number of validation examples $|\mathcal{E}_{test}|$,  validation time $T_{test}$, and rate of $\mathcal{E}_{test}$ examples {\em successfully validated} $\%\checkmark$) summarize the obtained results when validating the synthesized models in the $\mathcal{E}_{test}$ examples. An example is {\em successfully validated} iff the execution of the synthesized model, at the example pre-state, exactly produces the corresponding post-state of the example; all the synthesized action schemes generalize well to the new and larger examples in $\mathcal{E}_{test}$, validating every domain in less than 10s.

We also ran two experiments to compare our method with the performance of FAMA~\cite{aineto2019learning} at \strips{} domains, which are summarized as follows: (i), When we use our benchmarks to synthesize action models, FAMA meets the limitations of SAT-based approaches to handle hundreds of examples, getting stuck in the synthesis. (ii) When using the minimal learning sets from the FAMA benchmarks our synthesis performance gets close to FAMA, despite some of the synthesized  models fail to generalize because those minimal learning sets miss representative transitions. Last but not least, systems for learning planning action models, like FAMA~\cite{aineto2019learning}, or ARMS~\cite{yang2007learning}, assume that any state transition is modeled with a single \strips{} action so they fail to learn actions with {\em quantified effects}, models for cellular automata, or for the {\em pancake sorting} domain. 

\section{Conclusions}
We presented an innovative approach for synthesizing white-box models of transition systems as structured programs.  Our approach  synthesize compact models of state transitions that cannot be modeled by a single \strips{} action and that require {\em latent variables} missing
in the state representation; we showed that this kind of state transitions are produced by actions with {\em universally quantified conditional effects}, update rules of cellular automata, or vector manipulations like the {\scriptsize\tt flip} action from {\em pancake sorting}. We also showed that our approach can leverage different {\em target languages} (when available) to confine the solution space, without modifying the synthesis method. Our work is connected to the {\em heuristic search} approach to {\em generalized planning}~\cite{jimenez2019review,segovia2021generalized,segovia2022computing}; our research agenda is formalizing an effective heuristic search framework that subsumes both tasks. 

The computation of  properties and relations that generalize over world objects is studied since the early days of AI, with seminal works on the {\em blocksworld}~\cite{fikes1972learning,winston1975learning}. This family of computational problems is traditionally characterized by the use of FOL representations  and it is studied under different names and approaches, including ILP and {\em relational learning}~\cite{muggleton1992inductive,bergadano1996inductive,getoor2007probabilistic,de2008logical}. Full-observability of state transitions (the setting of this paper) is the most basic setting for this family of computational problems. We showed however that such setting is still challenging at domains where the number of state variables that are checked, or updated, is unbounded. We plan to extend our {\em structured programming} approach to address more challenging settings that consider {\em action discovery}~\cite{suarez2020strips,suarez2021online}, {\em partial observability}~\cite{amir2008learning}, {\em noisy examples}~\cite{mourao2012learning}, or {\em non-deterministic models}~\cite{pasula2007learning}, as it has been done in the {\em automated planning}~\cite{jimenez2012review,arora2018review,aineto2022comprehensive} and the ILP literature~\cite{raedt2016statistical}. 

\section*{Acknowledgments}
\begin{footnotesize}
This work is supported by the H2020 {\sc AIPlan4EU} project \#101016442. Javier Segovia-Aguas is  funded by TAILOR, AGAUR SGR and the Spanish grant PID2019-108141GB-I00. Jonathan Ferrer-Mestres is supported by the CSIRO MLAI Future Science Platform. Sergio Jiménez is funded by the Spanish MINECO project PID2021-127647NB-C22. 
\end{footnotesize}
\bibliography{ecai}
\end{document}